\documentstyle[12pt] {article}
\begin{document} 
 
\title{Pair Production of Open Strings - 
Relativistic versus Dissipative Dynamics}

\author{Ciprian  Acatrinei$^{1}$\thanks{E-mail acatrine@sissa.it; 
 on leave from 
\it{Institute of Atomic Physics - P.O. Box MG-6, 76900 Bucharest, Romania}}  
 and Roberto Iengo$^{2}$\thanks{E-mail iengo@he.sissa.it} \\
\\
{\it $^{1}$International School for Advanced Studies } \\
{\it Via Beirut 2-4, 34014 Trieste -Italy   }
\\
{\it ${^2}$International School for Advanced Studies and INFN} \\
{\it Sezione di Trieste, Via Beirut 2-4, 34014 Trieste -Italy  }}

\maketitle

\vskip1cm

\begin{abstract}
We study the pair production of open strings in 
constant external electric fields, using a general 
framework which encodes both relativistic string theory 
and generic linearly extended systems as well.
In the relativistically invariant case we recover 
previous results, both for pair production
and for the effective Born-Infeld action.
We then derive a non-relativistic limit - where the 
propagation velocity along the string is much smaller 
than the velocity of light - obtaining quantum dissipation. 
We calculate the pair nucleation rate for this case,
which could be relevant for applications. 
\end{abstract}

\centerline{\large \bf 1. Introduction and Summary}

\vskip0.5cm

The dynamics of quantum systems in the 
presence of external fields
has been of interest since the early 
days of quantum mechanics. 
Even later, when a quantum description 
of gauge fields became available,
such systems remained appealing,
basically because they allowed insigths 
into some nonperturbative physics.
One famous example is the Schwinger 
calculation \cite{schwinger} of the rate of 
production of electron-positron pairs out 
of the vacuum, in the presence
of a constant electric field. This has been achieved 
by computing a one-loop effective action
in the presence of a background.
 
In recent years, the dynamics of (real or virtual) strings in
electromagnetic backgrounds - coupled to the ends of
an open string - has also been studied. In particular, the
Born-Infeld \cite{bi} effective action for the external field has been
obtained in \cite{ft,acny},  
whereas the possibility of string pair creation through  
the Schwinger mechanism has been discussed in  refs. \cite{burgess,bp}.
The string vacuum also develops
a 'classical' instability \cite{burgess,nesterenko}, due to the
finite extension of the strings.
It has been shown that there is an interesting relation 
between string theory (ST) and quantum dissipation \cite{ct}.
This opens the possibility of studying the generalization of the 
Schwinger mechanism in the context of dissipative dynamics,
by means of ST methods.
Pair creation within dissipative dynamics has been investigated in 
ref.\cite{ij1}, related to possible applications to vortex pair nucleation
in a superconductor. There, quantum dissipation has been introduced by means of 
the Caldeira-Leggett (CL) formalism \cite{cl}.
Subsequently in \cite{ij2}, the relation between the CL
formalism and the dynamics of linearly extended system has
been explicitely worked out, in the context of non-relativistic
pair nucleation. This was obtained by integrating 
out the (free) degrees of freedom along the string, 
the resulting effective action for the end-point (where the 
external electric field couples) 
turning out to be of dissipative type, in some infinite string length limit.
For the standard relativistic string this limit, 
which corresponds to the case where the string worldsheet is
a disk, gives the Born-Infeld action for the external field. 
In order to get pair creation in that case, one has to consider a 
\emph{non-relativistic} version of ST.
Thus, pair creation within quantum dissipation appears to require
some non-relativistic limit, although the Schwinger mechanism
describes the antiparticle member of the pair as a particle moving
backward in time.

We wish to present here a systematic treatment of the
above problems, based on a generalization of the 
(bosonic) string action, which will allow us
to derive in a unified fashion:  

a) the above known results for 
the relativistic string, namely the rate of pair 
production in external fields for the case of an annulus 
world-sheet, as well as the Born-Infeld effective 
action obtained when the world-sheet shrinks to a disk ; 

b) the dissipative dynamics  
of the string end-point (which is meant to represent a physical
particle, say a pointlike vortex, coupled to the external field ) and
the pair nucleation rate in a nonrelativistic context.

We provide a \emph{continuous} 
interpolation between the results for relativistic 
string pair production studied in \cite{burgess,bp} and the 
dissipation-dominated  nucleation of refs \cite{ij1,ij2}. 
This is done by introducing a physical parameter, 
having the meaning of a velocity, 
which is allowed to be much smaller  than the speed of light.

Our method is reliable for such different situations 
because it is field theoretical in essence: 
we study
by path integral methods the dynamics of an 
oscillating linearly extended object. 
Actually, for standard relativistic string theory the results are 
valid only in the critical dimension,
whereas for the non-relativistic case they can apply to any 
space-time dimensionality. 
In this paper we focus on a (3+1) dimensional string 
theory, as demanded by possible applications  
to situations at an accessible energy.
At non critical dimensionality, of course, we have to face 
noncanonical values for the zero-point energy 
(coming from the normal ordering of the string 
Hamiltonian), which we will anyhow reabsorb into a 
renormalization of some rest energy.
In any case, we observe that the basic result for 
the Schwinger-like pair creation does not 
crucially depend on the string dimensionality. 
As every WKB tunnel effect, the nucleation rate 
is given by the product of an exponential factor 
and a mildly varying prefactor.
The exponent is not affected by the space-time
dimensionality and by the possible inclusion
of the fermionic degrees
of freedom of the supersymmetric version of the theory,
but it crucially depends on the type of dynamics, i.e.
whether it is dissipative or not. 
Namely, for small electric field $E$, it is
proportional to $-1/E$ in the usual
nondissipative dynamics, including the case of
the Schwinger mechanism for relativistic strings, 
whereas it goes like $-1/E^2$ for the dissipative case.

Having in mind the nonrelativistic application mentioned above
(say pair nucleation in a planar system),
our computations will include the contribution 
to the prefactor of (2+1) coordinates. The third one can be regarded as 
parametrizing the string extension 
(see fig.1, where the electric field is also indicated), and does not
contain physical degrees of freedom.  

\begin{picture}(405,150)(20,10)
\put(30,50){\line(1,0){200}}
\put(78,74){\line(1,0){200}}
\put(30,90){\line(1,0){200}}
\put(78,114){\line(1,0){200}}
\put(30,50){\line(2,1){48}}
\put(230,50){\line(2,1){48}}
\put(30,90){\line(2,1){48}}
\put(230,90){\line(2,1){48}}
\put(350,100){\vector(1,0){50}}
\put(350,100){\vector(-2,-1){40}}
\put(350,100){\vector(0,-1){50}}
\put(340,120){\vector(-2,-1){40}}
\put(330,120){E}
\put(295,75){$X_{1}$}
\put(400,90){$X_{2}$}
\put(360,55){$X_{3}$}
\qbezier(140,102)(136,95)(145,85)
\qbezier(145,85)(156,70)(150,61)
\put(140,102){\circle*{2}}
\put(150,61){\circle*{2}}
\put(180,30){fig.1  Stretched string}
\end{picture}

The plan of this paper is as follows. In section 2 
we briefly review the Schwinger computation of the 
vacuum decay rate, adapted to our case.
In section 3 we introduce a general form of the string 
action, which allows the description of both 
the usual relativistic string and the nonrelativistic case as well.
The later one will arise by taking a suitable limit 
for our parameters. We clarify that limit, 
explaining its significance. 
In section 4 we work out our calculations, which will 
be made with the general action. 
Here we evaluate - by path integral methods - the partition 
function in the presence of the external electric field coupled to
the end point of the open string.    
The appearence of singularities resulting from the path integral, namely  
poles in the worldsheet modulus, signals the possibility 
of pair production.
Relativistic and dissipative dynamics will be just
particular cases of this unifying picture.
In section 5 we use - for the same calculation - the 
'shorcut' provided by the boundary state 
formalism (BSF) [11], where the boudary conditions containing the electric
field are explicitely implemented.  We get in fact the same result, with a suitable
normalization of the boundary state.
We can say that the BSF provides a quick way to see 
what happens, but to get the proper normalization 
we have to compare with the path integration.
Section 6 is dedicated to the discussion of 
the pole singularity in various limits. We also show how to get 
the pole from a zero-mode of the action, corresponding to a solution of the
classical equations of motion with proper boundary conditions.
We compare our results in the relativistic limit 
with previous work \cite{ft,bp} in section 7. Indeed, 
we check that going to 26 dimensions
and including the ghosts, one gets the same rate of 
pair production as in \cite{bp}. 
The Born-Infeld action arises as in ref \cite{ft}, 
in the limit in which the world-sheet has only one boundary.
In section 8 we proceed by calculating the
rate of pair creation in the dissipative limit,  
ending with a few comments on the results.

\vskip0.5cm

\centerline{\large \bf 2. General setting}

\vskip0.5cm

Our final aim is to evaluate the probability of 
vacuum decay through pair production,
in an external constant electric field, in a 
bosonic string theory or a field theory. 
So, let us review briefly
the way to compute it. The zero temperature vacuum 
free energy in presence of an external 
constant electric field $E$ is the logarithm of:

$$
e^{-i W(E)_{(vac)} }=<0|e^{-i\hat{H}\times (time) }|0> 
\equiv <0|0>_{E}\equiv Z.
$$
For a static field $W_{vac}(E)={\cal E}_{vac}(E) 
\times (time)$, where $(time)$ is the 
total time interval.

We consider the vacuum fluctuations due to 
creation and annihilation of open strings. 
The zero temperature
vacuum free energy can be expressed by the 
following formula [1]
\begin{equation}
W_{vac}=\int_{0}^{\infty}\frac{dt}{t} 
Tr e^{-t \hat{H}_{string}}
\end{equation}
and the trace is evaluated by means of a 
(suitably normalised) path integral
$$
Tr e^{-t \hat{H}_{string}}=\int DX_{0}D\vec{X} e^{-S(X_{0}, \vec{X}, E)}.
$$
The action $S(X_{0}, \vec{X}, E)$ 
(discussed in section 3) 
includes an electric field, 
which couples to one of the string's end-points.
Due to that coupling, the vacuum energy 
has an imaginary part
$$
{\cal E}_{vac}(E)=Re {\cal E}_{vac}(E)-i\frac{\Gamma}{2} 
$$
which provides us with an expression for 
the vacuum decay rate per unit volume

\begin{equation}
\frac{\Gamma}{V}=-\frac{2}{(time)V}Im\int_{0}^{\infty}\frac{dt}{t}
\int DX_{0}D\vec{X}e^{-S(X_{0},\vec{X},E)}. 
\end{equation}
Here $V$ is the  
space volume spanned by the coordinates on 
which we path-integrate in eq.(3). 
Now, the zero modes in the path
integral will give precisely the space-time 
volume, so we obtain a formula for 
the vacuum decay rate per unit area $\gamma=\frac{\Gamma}{V}$:

\begin{equation}
\gamma=-2 Im\int_{0}^{\infty}\frac{dt}{t}\int' 
DX_{0}D\vec{X}e^{-S(X_{0},\vec{X},E)}. 
\end{equation}
The prime means that we have factored out the 
zero mode part of the action. 

The path integral factorizes into a free one (along $X_{2}$) 
and a second one including the free terms
along the axes $0$ and $1$, plus the interaction 
term involving $E$, which couples $X_{0}$ and $X_{1}$.

\vskip1.5cm

\centerline{\large \bf 3. Relativistic versus nonrelativistic limit }
\centerline{\large \bf  and quantum dissipation}
\vskip0.5cm

We wish now to explain in more detail which situations 
we are going to adress, and to see how both
relativistic and dissipative nonrelativistic dynamics 
arise from our formalism.
Since we take all the space directions to be on equal 
footing and the electric field along $X_{1}$, the action is:
$$
 S=-\frac{\alpha}{2}\int_{0}^{t} d\tau \int^{l}_{0}
 d\sigma[(\frac{\partial X^{0}}{\partial \tau})^{2}
 +(\frac{\partial X^{0}}{\partial \sigma})^{2}]
   +\frac{\beta}{2}\int_{0}^{t} d\tau \int^{l}_{0}d\sigma
   [ (\frac{\partial \vec{X}}{\partial \tau})^{2}+v^{2}
   (\frac{\partial \vec{X}}{\partial \sigma})^{2}]
$$
\begin{equation}   
   -iE\int_{0}^{t} d\tau \left[X_{0}
   \frac{\partial X_{1}}{\partial \tau}\right]_{\sigma=l}.
\end{equation}
For $v=1$ and $\alpha =\beta$, eq (4) is the standard 
way of writing the worldsheet action in string theory. 
But we have in mind
possible applications to dissipative quantum dynamics 
in nonrelativistic situations, thus we consider also $v<1$ and, 
for generality, $\alpha \neq \beta$
(although the precise relation between $\alpha$ and $\beta$ 
will not be needed in the following). 

Now we rescale  by $\tau \rightarrow \alpha l \tau , 
\sigma \rightarrow \sigma l$, 
so that the limits of integration become: 
\begin{equation}
T=\frac{t}{\alpha l} \quad \quad \Delta\sigma=1 .    
\end{equation}
$T$ gets the  dimension of a length 
to the square, and allows 
the explicit connection 
with Schwinger's "proper time" formalism. 
The case of quantum dissipation for 
nonrelativistic nonrelativistic 
situations corresponds to the limit:
\begin{equation}
\alpha T>>1>> \alpha v T .   
\end{equation}
We see what that means: the oscillations 
of $X_{0}$ along $\sigma_{0}$ 
are much suppressed compared to those along
$T$. For $\vec{X}$ the reverse happens. 
So time ($X_{0}$) 
is "rigid", but space ($\vec{X}$) not at all. 
We will compute the decay rate in these limits,
while keeping $\beta v$ fixed. $\alpha$ 
will set the scale. 

We now specialize to the (2+1)-dimensional case, 
that is $\vec{X}=(X_{1},X_{2},X_{3})$, where $X_{1,2}$ 
describe 
the physical transverse oscillations of the string, 
whereas we take $X_{3}=\sigma b$, with $b$ the 
intermembrane distance (see fig.1). 
Thus,  $X_{3}$ contributes to the 
action with the term:

\begin{equation}
 \int_{0}^{ T} d\tau \int^{1}_{0}d\sigma\frac{1}{2}
 \alpha\beta v^{2}(\frac{\partial b\sigma}{\partial \sigma})^{2}=
\frac{1}{2} T \alpha \beta v^{2}b^{2}. 
\end{equation}
The partition function will contain as a factor the 
exponential of minus this term, and (7) will play a 
role similar to the square rest mass
of a string.
Thus, we identify 
\begin{equation}
{\cal E}_{0}=\sqrt{\frac{1}{2}\alpha\beta v^{2}b^{2}}
\end{equation}
as being the rest energy of a string.

The physics of the situation is quite transparent, 
and shows clearly why this limit 
has to be called 'nonrelativistic'.
We have already seen that in the free action for $X_{0}$, 
the oscillations along $\tau$ (the 'kinetic term') 
dominate over those along $\sigma$.
Along the spatial directions the opposite happens. 
This is due just to $\alpha >>\frac{1}{T}\sim {\cal E}_{0}^{2}$ 
and is similar to what happens
in string theory when $\alpha' \sim\frac{1}{l_{S}^{2}}>>{\cal E}^{2}$, 
where ${\cal E}$ is the available energy. 
Then, the stringy massive modes are frozen,
in the same way in which the oscillations of $X_{0}$ 
along $\sigma$ are frozen in our problem.
Now, since $\tau$ has the dimensionality of time 
to the square, and the energy scale of our 
problem is ${\cal E}_{0}$, the natural time scale 
will be
$time = {\cal E}_{0} \tau$. In consequence, we have 

\begin{equation}
d(time)={\cal E}_{0} d\tau \quad \quad d(X_{3})=b d\sigma.
\end{equation}
Using the previous equation in the free action, 
we obtain the velocities with which $X_{0}$ 
and $X_{1,2}$ signals propagate, respectively,:
\begin{equation}
v^{2}_{prop}(X_{0})=\frac{\alpha}{\beta}\frac{1}{v^{2}}  
\quad \quad v^{2}_{prop}(X_{1})=\frac{\alpha}{\beta}.
\end{equation}
Furthermore, for our system the length scale is given 
by $b$, whereas a typical time scale is given by ${\cal E}_{0}^{-1}$. 
Thus a typical speed is of the order of

\begin{equation}
v_{typical}=b{\cal E}_{0}.
\end{equation}
It is now easy to see that the relation between 
these three velocities, thanks to the limit (6), 
is the following:
\begin{equation}
v_{prop}(X_{0})>>v_{typical}>>v_{prop}(X_{1,2}).
\end{equation}
The first inequality means that the propagation 
of time excitations is practically instantaneous 
(the time is Galilean) - a nonrelativistic situation. 
The second one says that the propagation of 
excitations of space-like coordinates is very slow compared 
to the typical velocity - 
or that strings are very long. This corresponds 
to pure dissipative dynamics, since in this limit 
one obtains the Caldeira-Leggett action (\cite{ct,ij2}).

We wish to see qualitatively why this nonrelativistic limit
is related to the Caldeira-Leggett type 
quantum dissipative dynamics. In \cite{cl}, dissipative 
dynamics was obtained by integrating out a thermal bath 
made of oscillators, on which a spectral condition has been 
imposed. The way we can reobtain
a "thermal bath" is very simple: we just 
rewrite $X(\sigma , \tau )$ as $X_{\sigma}(\tau )$. 
Then, we see that only $X_{1,2}$ depend
on $\sigma$, so the integration over a 
thermal bath is replaced here 
by path integrating over $\sigma$. The finite spatial 
extension of the strings amounts to an infinity 
of harmonic oscillators which can form a suitable bath. 
On the other hand, $X_{0}$ is independent
(in the nonrelativistic limit!) of $\sigma$, 
so the time is singled out as a good coordinate for a 
pointlike object from the beginning.
This is what we need, since we want 
to have the same time coordinate along the string to 
obtain nonrelativistic quantum mechanics,
whereas integrating out the spatial 
coordinates will provide us with a 
dissipative dynamics for the point 
particle which remains after \cite{ct,ij2}.
In CL language, time is already a macroscopical cordinate, 
whereas the space-like coordinates of the string's charged 
end are not; they can be made so only if accompanied
by a dissipative term. What is remarkable is the fact 
that the CL "spectral condition" is not needed. The string 
seems to \emph{automatically} provide such a constraint.
Of course, \emph{some} constraint is to be expected because 
all the dynamics is encoded in a continuum Lagrangian with 
fewer parameters than the 'many-body' CL one (for various 
approaches, see \cite{ct,ij2,cl}).

\vskip1.5cm

\centerline{\large \bf 4. Path integral evaluation}

\vskip0.5cm

{\bf 4.1 Free case ($E=0$)}

For completeness, but also to establish its full physical 
interpretation, we first evaluate the free string
 - properly normalized - partition function. 
We start from the action for a generic 
uncoupled coordinate $X$:
\begin{equation}
 S=\int_{0}^{t} d\tau \int_{0}^{l} 
 d\sigma [\frac{\alpha}{2}(\frac
{\partial X}{\partial \tau})^{2}+\frac{\alpha v^{2}}{2}
(\frac{\partial X}{\partial \sigma})^{2}]. 
\end{equation}
We take the  boundary conditions 
to be periodic along 
$\tau$: $ X(t+\tau,\sigma)=X(\tau,\sigma)$ 
and Neumann along $\sigma$ :
  $\frac{\partial X}{\partial \sigma}|_{\sigma=0,l} =0   $.
Of course, if we want to consider the time-like 
direction $X_{0}$, then $v=1$. 
The boundary conditions allow 
us to write

\begin{equation}
X(\tau,\sigma)=\sum_{n\in Z}\sum_{k\in N}X_{nk}cos(k\pi\frac{\sigma}{l})
exp(2\pi in\frac{\tau}{t}) 
\end{equation}
such that the action becomes

$$S=\alpha \frac{tl}{2}\left(\sum_{n>0,k>0}|X_{nk}|^{2}
[\frac{4\pi^{2}}{t^{2}}n^{2}+v^{2}\frac{\pi^{2}}{l^{2}}k^{2}] 
    +\sum_{n>0,k=0}|X_{n0}|^{2}[2\frac{4\pi^{2}}{t^{2}}n^{2}]
   +\sum_{n=0,k>0}X_{0k}^{2}[\frac{v^{2}}{2}\frac{\pi^{2}}{l^{2}}k^{2}] \right) $$
We drop the zero-mode ($n=0, k=0$) in the path integral, 
since it gives just the volume of space-time, 
and we wish to compute the amplitude 
per unit time and volume. 
Then we have to evaluate 

%\begin{displaymath}
\begin{eqnarray*}
\int DX e^{-S} & = & \prod_{k>0}\frac{1}{\sqrt{\frac{\pi t}{4l}k^{2}\alpha v^{2}}} 
\prod_{n>0}\frac{1}{\frac{4\pi l}{t}n^{2}\alpha} 
\prod_{k>0}\prod_{n>0}\frac{1}{\alpha\frac{lt}{2\pi}
( \frac{4\pi^{2}}{t^{2}} n^{2}+ \frac{\pi^{2}}{l^{2}}k^{2})} \\
& &     \\
& = & \sqrt{\frac{\alpha l}{2\pi t}}e^{-\pi\frac{tv}{2l}\sum_{k>0}k}
\prod_{k>0}\frac{1}{1-e^{-2\pi k\frac{tv}{2l}}}. 
\end{eqnarray*}
%\end{displaymath}
We have used - only for the index 
$n$, corresponding to the 
Fourier transform along $\tau$ - 
the free particle normalization
$$ \prod_{n>0}\frac{1}{4\pi}\frac{t}{m}
\frac{1}{n^{2}}=\sqrt{\frac{m}{2\pi t}}.$$
as well as the Euler factorization of 
$\frac{\sinh{x}}{x}$:

$$
\prod_{n=1}^{\infty}(1+\frac{x^{2}}{n^{2}})
=\frac{\sinh{\pi x}}{\pi x}.
$$
Now we make use of the transformation 
properties of the Dedekind eta function
\begin{equation}
\eta(x)=e^{i\pi \frac{x}{12}}\prod_{1}^{\infty}(1-e^{2\pi inx})=
\frac{1}{\sqrt{-ix}}\eta(-\frac{1}{x})
\end{equation}
and get at the end:

\begin{displaymath}
\int DX e^{-S}=\sqrt{\frac{\alpha v}
{4\pi}}e^{-\frac{\pi}{2}\frac{vt}{l}\sum_{k>0}k} 
e^{-\frac{\pi}{12}\frac{vt}{2l}} e^{\frac{\pi}{12}\frac{2l}{vt}}
\prod_{n>1}\frac{1}{1-e^{-4\pi n \frac{l}{vt}}}.
\end{displaymath}

The factor
$e^{-\frac{\pi}{2}\frac{vt}{l}\sum_{k>0}k} 
e^{-\frac{\pi}{12}(\frac{vt}{2l})}$
becomes one if we use the Riemann 
$\zeta$-function regularization 
$ \sum_{k=1}^{\infty}=lim_{s \rightarrow 1}\zeta(-s) = -\frac{1}{12}$,
as in string theory. 
The remaining exponential term is important. 
We will see that it amounts to 
normal order the Hamiltonian in the boundary 
state formalism, 
and it gives a contribution which we can 
interpret in a thermodynamical context.

This is so because, if we make the 
identification $ 1/t = T = \beta^{-1} $, 
were $T$ means temperature now
($\tau$ has been already Euclidean), we 
can interpret the path integral as 
a partition function
of a system in thermal equilibrium:

$$ \int DXe^{-S}=Z_{0}=\sum_{n}e^{-\beta E_{n}}
=e^{-\beta F} =e^{-W} .$$
The specific heat of this system \cite{is} is then 
given by ($U=\frac{\partial W}{\partial \beta}$):

\begin{displaymath}
 C=\left(\frac{\partial U}{\partial T}\right)_{V=cst} 
= \frac{\pi}{3}T\frac{l}{v}, 
\end{displaymath}
which is positive (in our case  
$W=-\frac{\pi}{12}\frac{2l}{vt}$) and raises linearly
with the temperature. This indicates that 
\begin{equation}
c=\frac{\partial C}{\partial T}=\frac{\pi}{3}\frac{l}{v}
\end{equation}
is a  temperature independent physical parameter, 
which could in 
principle be measured. It will appear in our result
for string pair production 
in the nonrelativistic case, when 
the string could be identified with a vortex line.  

We may say that a system with an 
infinity of degrees of freedom, like
a string, can be characterized 
by a peculiar specific heat 
in a thermodynamical context. 
Thus, our final result is

\begin{equation}
\int DX e^{-S}=\sqrt{\frac{\alpha v}{4\pi}}e^{\frac{1}{2t}c}
\prod_{n\geq 1}\frac{1}{1-e^{-4\pi n \frac{l}{vt}}}.
\end{equation}

\vskip0.5cm

{\bf 4.2 Interacting case ($E\neq 0$)}

\vskip0.2cm

The normalization we have obtained 
up to now was for the free case.
If we switch on the electric field, 
however, we could get a normalization
factor depending on the magnitude 
of $E$, so that we can not 
rely on what we have just learnt 
in the free case. In order to see 
what happens in general, 
with correct normalizations included,
we evaluate the path integral 
for the interacting case.

The interaction term is quadratic, so that 
we could hope to eliminate it by a
suitable linear transformation of $X_{0}$ 
and $X_{1}$. This does not work directly,
since our interaction is a boundary term, 
not a bulk term.

Our strategy will be to decompose in Fourier 
modes, and evaluate the determinant
in momentum space. In order to evaluate:

$$
Z=\int DX_{0} DX_{1} e^{-[\int_{0}^{t} d\tau \int^{l}_{0}d\sigma [-\frac{\alpha}{2}
(\frac{\partial X^{0}}{\partial \tau})^{2}-\frac{\alpha}{2}
(\frac{\partial X^{0}}{\partial \sigma})^{2}+\frac{\beta}{2}
 (\frac{\partial X^{1}}{\partial \tau})^{2}+\frac{\beta v^{2}}{2}
(\frac{\partial X^{1}}{\partial \sigma})^{2}]
-i\int_{0}^{t} d\tau [EX^{0}\frac{\partial X^{1}}{\partial \tau}]_{\sigma =l}]}
$$
we develop in an arbitrary basis, 
independent of the interaction, 
as in the free case:

$$ X(\tau,\sigma)=\sum_{n\in Z}\sum_{k\in N}X_{nk}
cos(k\pi\frac{\sigma}{l})exp(2\pi in\frac{\tau}{t}) .$$
Then $S$ becomes :
$$
S=S(0)+\sum_{n>0}S(n),
$$
with $$ S(0)=-\frac{tl}{2} \sum_{k>0}[(X^{0}_{0k})^{2} 
\frac{\alpha}{2}\frac{\pi^{2}}{l^{2}}k^{2}-
(X^{1}_{0k})^{2} \frac{\beta v^{2}}{2}\frac{\pi^{2}}{l^{2}}k^{2}]
$$
and
$$
S(n>0)=  -\frac{tl}{2}[|X^{0}_{n0}|^{2} 4[\frac{\alpha}{2}\frac{4\pi^{2}}{t^{2}}n^{2}]-
|X^{1}_{n0}|^{2} 4[\frac{\beta}{2}\frac{4\pi^{2}}{t^{2}}n^{2}]]   +{\bf X^{\dagger} }A{\bf X}.
$$
The term ${\bf X^{\dagger} }A{\bf X}$ 
encodes the modes which 
couple through the electric field; 
we use the notation:

$$ {\bf X^{\dagger}}=( X^{0}_{-n,1} \dots X^{0}_{-n,k} \dots \quad  
X^{1}_{-n,1} \dots X^{1}_{-n,k} \dots) $$
and
\begin{displaymath}
A = \left ( 
\begin{array}{ccccccccccccc}
a_{1} & 0 & 0 & \cdot & \cdot &  &  D & D & D & \cdot & \cdot &  \\
0 & a_{2} & 0 & \cdot & \cdot &  &  D & D & D & \cdot & \cdot &   \\
\cdot & \cdot & \cdot & \cdot &  &  & \cdot & \cdot & \cdot & \cdot &  &  \\
-D & -D & -D  & \cdot & \cdot &  &  b_{1} & 0 & 0 & \cdot & \cdot &  \\
-D & -D & -D  & \cdot & \cdot &  &  0 & b_{2} & 0 & \cdot & \cdot &  \\
\cdot & \cdot & \cdot & \cdot &  &  & \cdot & \cdot & \cdot & \cdot &  &    
\end{array}               
\right ).
\end{displaymath}
Here the 'coupling term' is $ D=-2\pi neE$, whereas 
$a_{k>0}=\alpha\frac{tl}{2}
(\frac{4\pi^{2}}{t^{2}}n^{2}+\frac{\pi^{2}}{l^{2}}k^{2}) $ 
and similarly for $b_{k}$, 
with $\alpha,\alpha \rightarrow \beta,\beta v^{2}$.
It turns out that (after some algebra 
and a proof by induction for any $n$):  
$$
det(A)=\prod_{i=1}^{n} a_{i} \prod_{j=1}^{n} b_{j} 
\left[1+D^{2}(\sum_{i=1}^{n}\frac{1}{a_{i}})
(\sum_{j=1}^{n}\frac{1}{b_{j}})\right].
$$
Using that, together with the identity

$$ \sum_{k=1}^{\infty}\frac{1}{a^{2}+k^{2}}
=\frac{1}{2}[\frac{\pi}{a}cth(\pi a)-\frac{1}{a^{2}}],  $$
our path integral becomes

\begin{equation}
\begin{array}{rl}

 Z & =(Z_{free})\prod_{n\geq 1}\left(1- \frac{E^{2}}{\alpha \beta v} 
cth(2\pi n \frac{l}{t}) cth(2\pi n \frac{l}{vt})  \right)^{-1}  \\
  &                 \\
  & = [1-\frac{E^{2}}{\alpha \beta v}]^{\frac{1}{2}}
 \sqrt{\frac{\alpha \beta v}{(4\pi )^{2}}} 
 e^{\frac{\pi}{6}\frac{l}{t}(1+\frac{1}{v})}  \\
  & \\
  &  \quad  \times \prod_{n\geq 1}\frac{1}
 {[1+e^{-4\pi n\frac{l}{t}(1+\frac{1}{v})}
 -\frac { 1+\frac{E^{2}}{\alpha \beta v} } {1-\frac{E^{2}}{\alpha \beta v} } 
 ( e^{ -4\pi n\frac{l}{t}} + e^{ -4\pi n\frac{l}{vt}} ) ]} .
\end{array}
\end{equation}
We have $\zeta$-function regularised the divergent sum $\sum_{k>0} 1
=lim_{s\rightarrow 0}\zeta (s) = -\frac{1}{2}$.

\vskip1.5cm

\centerline{\large \bf 5. Boundary State Formalism (BSF) evaluation}

\vskip0.5cm

We are going to repeat the former computation, this 
time through the boundary state formalism, that is, implementing 
in an operatorial way the boundary conditions in presence of the
electric field. A review is presented in Appendix A. 
The result eq.~(41) can be written as:
\begin{equation}
\int DX^{0}(\sigma ,t)DX^{1}(\sigma ,t) 
e^{-S}=N(E)<B_E(l)|B_{0}>=N(E)<B_{E}|e^{-lH}|B_{0}>,
\end{equation}
where $N(E)$ is a normalization factor to be obtained later by comparing 
with the path integral result (18). 
The boundary state $|B_{E}>$ is obtained in eq.(40) of Appendix A. 
Using the notation    

$$
\frac{E}{\alpha}=A \quad \quad \quad  2\pi \frac{l}{t}=l_{0}   
$$
$$
\frac{E}{\beta v}=B \quad \quad \quad  2\pi \frac{l}{vt}=l_{1}   
$$
which will simplify the writing, we obtain:

\begin{displaymath}
<B_{E}(l)|B_{0}(0)>=<0|exp\sum_{n \geq 1}[\frac{AB+1}{AB-1}
(X_{n}^{1}\tilde{X}_{n}^{1}e^{-2nl_{1}}-X_{n}^{0}\tilde{X}_{n}^{0}
e^{-2nl_{0}}) 
\end{displaymath}
\begin{equation}
+\frac{2}{AB-1} (AX_{n}^{0}\tilde{X}_{n}^{1}-BX_{n}^{1}
 \tilde{X}_{n}^{0})e^{-nl_{0}}e^{-nl_{1}}]
exp\sum_{n \geq 1}[X_{-n}^{0}\tilde{X}_{-n}^{1}-X_{-n}^{1}\tilde{X}_{-n}^{0}]|0> 
\end{equation}
where $l$ plays now the role of 'time'. The former expression 
will be given by an infinite product of terms of the form

\begin{displaymath}
<0|e^{\kappa b\tilde{b}}e^{\lambda a\tilde{a}}e^{\mu a\tilde{b}}e^{\nu b\tilde{a}}
e^{\rho a^{\dagger}\tilde{a}^{\dagger}}e^{\sigma b^{\dagger}\tilde{b}^{\dagger}}|0>, 
\end{displaymath}
the operators $a,a^{\dagger}$ and $b,b^{\dagger}$ 
satisfying the usual
harmonic oscillator commutation relations.
Using the formula (proved in Appendix B)
\begin{equation}
<0|e^{\kappa b\tilde{b}}e^{\lambda a\tilde{a}}e^{\mu a\tilde{b}}e^{\nu b\tilde{a}}
e^{\rho a^{\dagger}\tilde{a}^{\dagger}}e^{\sigma b^{\dagger}\tilde{b}^{\dagger}}|0>
=\frac{1}{(1-\kappa\sigma)(1-\lambda\rho)-\mu\nu\rho\sigma}
\end{equation}
and substituting back , we get 

$$
Z=N(E)<B_{E}(l)|B_{0}>= N(E) e^{\frac{\pi }{12}\frac{2l}{t}} 
e^{\frac{\pi }{12}\frac{2l}{vt}}
$$
\begin{equation}
\times  \prod_{n>0} \frac{1}
 {1+e^{-4\pi n\frac{l}{t}(1+\frac{1}{v})}
 -\frac {1+\frac{E^{2}}{\alpha\beta v}} {1-\frac{E^{2}}{\alpha \beta v}} 
 (e^{ -4\pi n\frac{l}{t}}+e^{ -4\pi n\frac{l}{vt} })} .
\end{equation}
In order to reproduce the path integral 
result we have to take 
$N(E)=N_{0}\sqrt{1-\frac{E^{2}}{\alpha \beta v}}$,
$N_{0}=\frac{\sqrt{\alpha\beta v}}{4\pi}$.
We will see that the term $\sqrt{1-\frac{E^{2}}{\alpha \beta v}}$ 
 is going
to be just the Born-Infeld action, 
specialized for our case. 

\vskip0.5cm

It is interesting to point out the origin 
of the various factors.
If we have a closer look at the free string computation 
(15-16), we can see that 
$\frac{(\alpha \beta v)^{\frac{1}{2}}}{4\pi}$
comes from the zero modes along $\sigma$, i.e. it 
is related to the pointlike 
component of our object, not to its stringy excitations.
Also, it is the part of the free string partition 
function \emph{not} 
cancelled by the ghosts.  
The exponential term has already been interpreted 
as a specific heat 
contribution, characterizing the way energy is distributed
among the string modes (degrees of freedom). Finally, 
$\sqrt{1-\frac{E^{2}}{\alpha \beta v}}$ 
is a boundary term
(the electric field acts on the world sheet boundary, not 
on the bulk), and it has a long story. Its form will 
give rise - for the relativistic
string - to the Born-Infeld action. That factor has already 
been obtained in a variety of ways, for instance doing 
the path integral (after integrating out the bulk)
in configuration space [3], or through operatorial methods, 
developing in a Fourier basis adapted to the 
form of the interaction [8].

\vskip1.5cm

\centerline{\large \bf 6. Poles }

\vskip0.5cm

{\bf 6.1 General pole equation and particular limits}

\vskip0.5cm

>From now on, we switch to the notation of the rescaling (5). 
That simply amounts to set $l=1$ and replace $t=\alpha T$.
First, let us consider $ \alpha, \beta ,v $ arbitrary. 
In the general case (18,23) we have poles in $T$ whenever

\begin{equation}
\frac{E^{2}}{\alpha\beta v}=th 2\pi n\frac{1}{\alpha T} \cdot 
th 2\pi n\frac{1}{v\alpha T}.
\end{equation}
This is the general case from which, taking various limits, 
we obtain different physical situations.

$1)$ We can look at the case of the relativistic string
$ \alpha=\beta$, $v=1 $.
Poles are located, for a given $n$, at:

$$ \frac{E}{\alpha}=th2\pi n \frac{1}{\alpha T} $$
For either small $E$ or small $\frac{1}{\alpha T}$ we get

\begin{equation}
T= \frac{2\pi n}{E}.
\end{equation}
As we will see, this is the situation studied in \cite{bp}.

\vskip0.2cm

$2)$ A second possibility is $ \frac{1}{\alpha T}>>1 ,  
\frac{1}{v \alpha T}>>1 $,
either for the relativistic invariant case or in general.
In that case the pole is independent of $T$, and is 
located at the 'critical' value of $E$:

$$ \frac{E^{2}}{\alpha\beta v}=1 .$$
It corresponds to the case studied in \cite{ft}, were the 
Born-Infeld action was obtained.

$3)$ If we take the limit $\frac{1}{v\alpha T}>>1$ and
either $\frac{1}{\alpha T}<<1$ or $E$ small, we get

$$ T=\frac{2\pi n \beta v}{E^{2}}   .$$
We will see that this case corresponds to a nonrelativistic 
string, from which dissipative
point particle quantum mechanics can be obtained by 
integrating out the string degrees of freedom \cite{ij2}.
We will have more to say about this later.

\vskip0.5cm

{\bf 6.2 Another method}

We note that it is possible to obtain the expression 
for the poles also by solving the 
Euclidean equations of motion, subject to the boundary 
conditions established in Appendix A. 
Our expression for the poles (23) 
is just the consistency condition 
needed in order for the boundary conditions at $\sigma=0,1$ to be 
satisfied by both $X_{0}$ and $X_{1}$.

Taking

$$ \frac{\partial^{2}X_{n}^{0}}{\partial\sigma^{2}}=
\omega_{n}^{2} X_{n}^{0}  \quad \quad
\frac{\partial^{2}X_{n}^{1}}{\partial\sigma^{2}}=
\frac{1}{v^{2}}\omega_{n}^{2} X_{n}^{1} $$
we obtain ($\omega_{n}=\frac{2\pi n}{\alpha T}$):

$$
X_{n}^{0}(\sigma)=A_{0}ch\omega_{n}\sigma +B_{0}sh\omega_{n}\sigma \quad
X_{n}^{1}(\sigma)=A_{1}ch\frac{\omega_{n}}{v}\sigma +B_{1}sh\frac{\omega_{n}}{v}\sigma.
$$
Using $ \frac{\partial X_{n}^{0}}{\partial\sigma}|_{\sigma=0}=
        \frac{\partial X_{n}^{1}}{\partial\sigma}\mid_{\sigma=0}=0 \quad $
we get  $B_{0}=B_{1}=0$.
Finally, from the boundary conditions at $\sigma=l\equiv 1$ 
(see Appendix A) we get

$$ A_{0}\omega_{n}sh(\omega_{n})=\frac{\omega_{n}E}{\alpha}A_{1}ch(\frac{\omega_{n}}{v}) $$
$$ A_{1}\frac{\omega_{1}}{v}sh(\frac{\omega_{n}}{v})=\frac{\omega_{n}E}{\beta v}A_{0}ch(\omega_{n}).$$
For consistency then:
$$
\frac{E^{2}}{\alpha \beta v}=th(2\pi n\frac{1}{\alpha T})th(2\pi n\frac{1}{v\alpha T})
$$
which is our former pole condition. In fact, 
the possibility of 
having a classical solution for a particular $T$ for which
the boundary conditions at both ends can be satisfied implies 
that the (Euclidean) action of the corresponding
modes is zero. Thus the gaussian integration over 
those modes produces a singularity. 

\vskip1.5cm

\centerline{\large \bf 7. Comparison with previous results}

\vskip0.5cm

We wish now to compare the result we have obtained 
with previous ones.
First we remark that we can rewrite our partition 
function $Z$ in the form:
 
\begin{eqnarray*}
Z   & \sim & \prod_{n=1}^{\infty} \frac{1}{\frac{E^{2}}{\alpha\beta v}-
             th 2\pi n\frac{1}{\alpha T} th 2\pi n\frac{1}{v\alpha T} } \\
    &      & \\          
    & =    &   \prod_{n=1}^{\infty} det \left |
     \begin{array}{cc}
         th(2\pi n\frac{1}{\alpha T})  &  -\frac{E}{\alpha} \\
         \frac{E}{\beta v}            &  -th(2\pi n\frac{1}{v\alpha T})
     \end{array}  \right |^{-1} \\
    &      &   \\ 
    & =    &   \prod_{n=1}^{\infty} det \left (g^{\mu\nu} th(2\pi n\frac{1}{v_{\mu}\alpha_{\mu}T}) 
            -\frac{F_{\mu\nu}}{\alpha_{\mu}v_{\mu}} \right )^{-1}  ,\\
\end{eqnarray*}
where $\alpha_{\mu}, v_{\mu}$ are the string tension
 and propagation velocity for the coordinate $X_{\mu}$.
This makes clear the way the Born-Infeld action emerges: 
when the hyperbolic tangent above goes to one.  
Of course, we have proved the equality in the last line 
only for one electric field $F_{01}$ 
along the $X_{1}$ direction.
Nevertheless, the formula can be extended 
to a general constant $ F_{\mu\nu}$, 
the generalization requiring just to put 
an antisymmetric matrix into its block
diagonal form. This way of presentation 
is appropriate for studying 
various particular cases.

We made our calculations on a cylinder, of 
circumference $t$ and length $l$.
Through a conformal mapping, it can be
 transformed into an annulus, on which 
 some of the
previous calculations have been done. We 
remark that the 
limit $ l \rightarrow \infty $ (the present 
$\alpha T \rightarrow 0$) corresponds to the shrinking
of the interior circle of the annulus to zero 
radius, thus obtaining a disk.

\vskip0.3cm
{\bf 7.1 Relativistic string}  

If we take a relativistic string, $\alpha=\beta$ and $v=1$ , 
and keep $\alpha T$ finite, 
our results should be the same
as the ones of \cite{bp}. To check this, we rewrite 
our partition function $Z$, 
for $X_{0}$ and $X_{1}$ - see eq.(18), 
in terms of $\eta$ and $\Theta$ functions:

\begin{equation}
Z=-\frac{i}{2\pi}E
\frac{\eta(\tau=2i\frac{1}{\alpha T})}
{\Theta_{1}(u=-\frac{i}{2\pi}
ln(\frac{1+\frac{E}{\alpha}}{1-\frac{E}{\alpha}})|\tau)}.
\end{equation}
To prove that, put 
$ e^{2\pi i u}=e^{w}=\frac{1+\frac{E}{\alpha}}{1-\frac{E}{\alpha}} $ 
and 
$ q= e^{-4\pi \frac{1}{\alpha T}}
=e^{2\pi i \tau}$. Remembering that 
$$\Theta_{1}(u|\tau)=2q^{\frac{1}{8}} \sin{\pi u} 
\prod_{n=1}^{\infty}(1-q^{n})(1-e^{2\pi i u}q^{n})(1-e^{-2\pi i u}q^{n}) $$
we can rewrite our two-dimensional partition function as:

$$
Z=\sqrt{1-\frac{E^{2}}{\alpha^{2}}}\frac{\alpha}{4\pi}2 \sin{\pi u}
\frac{\eta(\tau=2i\frac{1}{\alpha T})}
{\Theta_{1}(u=-\frac{i}{2\pi}ln(\frac{1+\frac{E}{\alpha}}{1-\frac{E}{\alpha}})|\tau)}.
$$ 
Using the relationship between $u$ and $\frac{E^{2}}{\alpha^{2}}$ 
and the fact that $\sin{\pi u}=-i\sinh{\frac{w}{2}}$ we obtain (25).

The free one-dimensional partition function 
(along a space-like direction, $X_{2}$ say) 
becomes
$$
Z_{2}=\sqrt{\frac{\beta v}{4\pi}}\eta(\frac{2i}{\alpha T}).
$$

Now, if we take into account the other 24 
free dimension and the ghosts, 
and also use the way $\eta$ and $\Theta$ 
functions behave under modular transformations,
we obtain the full 26-dimensional bosonic partition function 

\begin{equation}
 Z_{26}=-\frac{i}{2\pi}E\frac{\alpha^{13}}{(4\pi)^{13}}\eta^{-21}(\frac{2i}{\alpha T})
\Theta^{-1}_{1}(u|\frac{2i}{\alpha T}).
\end{equation}

This is exactly the Bachas and Porrati amplitude for 
the case of an open bosonic string with a non-zero charge 
only at one end. Although a calculation involving superstrings 
would give rise to additional factors in the amplitude, it does 
not change the pole structure \cite{bp}, which is in fact given
by eq.(24). For this reason
we  restrict ourselves to the bosonic case. As discussed in ref \cite{bp},
this amplitude has poles in $T$, due to the zeroes of the $\theta$-function
$\Theta_{1}$. These poles induce the imaginary part in the r.h.s. of eq.(3) 
and thus the vacuum decay rate.  

\vskip0.3cm
{\bf 7.2 Born-Infeld action} 

Let us take the relativistic string case again, but now supplement 
it with the limit $\alpha T \rightarrow 0$ ($ l \rightarrow \infty $) 
- as we said, the annulus is shrinking to a disk.
Then, since $ th(\infty)=1$, we see that our 
expression for $Z$ reduces to the Born-Infeld action.
In this way we reobtain the result of \cite{ft}. 
In this case there is no pair production,
at least for $E<\alpha $. We remark that in our 
case the string has a charge only at one end, whereas
in \cite{ft} it had equal and opposite charges at the 
two ends, both coupled to the electric field.

We notice that it has been also observed in ref \cite{ij2} 
that the Born-Infeld action 
can be obtained by tracing out the string degrees 
of freedom in the relativistic case and 
the limit $ l \rightarrow \infty $.

In this limit, we have already seen 
that there are no poles (except 
for $E\rightarrow E_{critical}$),
hence no pair production.  
Indeed, in the limit $\alpha  T \rightarrow 0$ the 
26-dimensional partition function (26) reduces to
$$ \frac{\alpha^{13}}{(4\pi)^{13}}(e^{4\pi\frac{1}
{\alpha T}})\sqrt{1-\frac{E^{2}}{\alpha^{2}}} $$
which is (modulo a constant) a Born-Infeld action.

\vskip1.5cm

\centerline{\large \bf 8. Vacuum Decay Amplitude in a Dissipative Context}

\vskip0.5cm

We proceed now with the nonrelativistic case, 
where instead $v<<1$ will be a crucial condition.        
We put everything together to evaluate the decay rate , 
which is given by

$$
\gamma= -2Im\int_{0}^{\infty}\frac{dT}{T}\int DX_{0}DX_{1}DX_{2}e^{-S},
$$
multiplied by the rest mass factor (cf. section 3) coming from 
the fact that our strings are stretched along $X_{3}$.
We will work with the rescaled variable $T=\frac{t}{\alpha}.$  
We obtain the vacuum transition amplitude (over unit space and time, 
since we have already substracted the zero mode - cf. eq. (4)) 

\begin{displaymath}
 \gamma=-2Im\int_{0}^{\infty}\frac{dT}{T}\sqrt{\frac{\alpha(\beta v)^{2}}{(4\pi)^{3}}} 
  e^{\frac{\pi}{6}\frac{1}{\alpha T}(1+\frac{2}{v})}  e^{-\frac{1}{2}\alpha\beta v^{2}b^{2} T}
  \sqrt{1-\frac{e^{2} E^{2}}{\alpha\beta v}} 
\end{displaymath}
\begin{equation}  
 \times \prod_{n>0}  \frac{1}{[1-e^{-4\pi n\frac{1}{\alpha v  T}}]
 [1+e^{-4\pi n\frac{1}{\alpha T}(1+\frac{1}{v})}
-\frac{1+\frac{E^{2}}{\alpha\beta v}}{1-\frac{E^{2}}{\alpha\beta v}} 
(e^{-4\pi n\frac{1}{\alpha  T}}+e^{-4\pi n\frac{1}{\alpha v T}})]} . 
\end{equation}
In order to get the rate of pair production we 
focus on the imaginary part 
- due to the presence of poles - of 
the former expression. 
Taking the limits $ \alpha T>>1 $ 
and $\alpha vT<<1$
while keeping $\alpha$ and $\beta v$ 
fixed we remain with :

\begin{displaymath}
\gamma=  -2Im\int_{0}^{\infty}\frac{d T}{ T}
\sqrt{\frac{\alpha(\beta v)^{2}}{(4\pi)^{3}}} 
  e^{\frac{\pi}{6}\frac{1}{\alpha T}(1+\frac{2}{v})}  
  e^{-\frac{1}{2}\alpha\beta v^{2}b^{2} T}
  \sqrt{1-\frac{e^{2} E^{2}}{\alpha\beta v}} 
  \prod_{n>0} 
 \frac{1}{ [1-\frac{1+\frac{E^{2}}{\alpha\beta v}}
 {1-\frac{E^{2}}{\alpha\beta v}} 
(e^{-4\pi n\frac{1}{\alpha T}})]} 
\end{displaymath} 

We take into account only the first, dominant pole ($n=1$)
and evaluate the residue 
there using the expansion:

$$ \frac{1}{1-e^{\frac{2E^{2}}{\alpha\beta v}-4\pi\frac{1}{\alpha T}}}\simeq 
\frac{1}{-\frac{2E^{2}}{\alpha\beta v}+4\pi \frac{1}{\alpha T}}=
\frac{\frac{-\alpha\beta v}{2E^{2}} T}{T-2\pi \frac{\beta v}{E^{2}}} .$$
Thus we have a pole for $T=T_{P}=\frac{2\pi \beta v}{E^{2}}$.
Using now the identity $\frac{1}{x-i\epsilon}
=P(\frac{1}{x})+i\pi \delta(x)$ in order to obtain the imaginary part
of the $T$-integral in eq (27), we get the vacuum decay rate:

\begin{equation}
\gamma= \frac{1}{8\pi}\frac{\alpha(\beta v)^{\frac{3}{2}}}{E}
\sqrt{1-\frac{E^{2}}{\alpha\beta v}}
e^{\frac{E^{2}}{6\alpha\beta v^{2}}} 
e^{\frac{\pi^{2}}{12}\frac{\alpha\beta v}{E^{2}}}
e^{-\pi \alpha\beta^{2}v^{3}\frac{b^{2}}{E^{2}}}.
\end{equation}
We remark that we have used the transformation properties of the Dedekind $\eta$-function, eq (15).

We reinterpret the exponents in terms of physical quantities.
The quantum dissipation coefficient is $\eta=\beta v$ (see \cite{cl, ij2}).
The rest energy of the nucleated object is 
${\cal E}_{0}=\sqrt{\frac{1}{2}\alpha\beta v^{2}b^{2}}$
(see eq.(8)), thus 
$\pi \alpha\beta^{2}v^{3}\frac{b^{2}}{E^{2}}={\cal E}_{0}^{2}\cdot T_{P}$.
Further, we write 
$$
\frac{\pi^{2}}{12}\frac{\alpha\beta v}{E^{2}} 
\equiv -\Delta{\cal E}^{2}\cdot T_{P}
$$ 
(with $\Delta{\cal E}^{2}=-\frac{\pi \alpha}{24}$), 
reabsorbing it into a redefinition of the rest energy: 
${\cal E}^{2}={\cal E}_{0}^{2}+\Delta{\cal E}^{2}$.
Concerning the first term, we rewrite it
in terms of the temperature derivative of the specific heat,
$c=\frac{\pi}{3\alpha v}$ (see eq.(16)):
$$
\frac{E^{2}}{6\alpha\beta v^{2}}=c\frac{E^{2}}{4\pi \eta} .
$$
In physical applications, both ${\cal E}^{2}$ and $c$ will be taken as physical
parameters to be determined experimentally.  

We note one further point : in order to 
fix the normalization of the 
electric field $E$ we have to reember 
that in the Schwinger method
the space-time trajectory of a particle 
is described by a path integral action 
$S$ which includes a term $\frac{1}{4}\int_{0}^{ T}
d\tau(\frac{\partial X_{0}}{\partial \tau})^{2}$. 
Thus in eq.(28) we have to rescale $X_{0}$, 
which ultimately implies rescaling 
$E\rightarrow E\sqrt{2}$. Finally we get the 
decay rate :
\begin{equation}
\gamma= \frac{1}{8\sqrt{2}\pi}\frac{\alpha(\eta)^{\frac{3}{2}}}{E}
e^{\frac{E^{2}}{2\pi \eta}c} e^{-\pi \eta\frac{{\cal E}_{0}^{2}}{E^{2}}}.
\end{equation}
We note that our result has the 
same form as equation (13) of the second reference in \cite{ij1}. 
It is remarkable that while there a cut-off 
on the frequency for which dissipation occurs has
been introduced by hand, here the way we calculate 
the path-integral starting with an underlying 
string theory takes care of everything.

We now discuss in which conditions a production 
rate should be observable.
For that, we would need  
$T {\cal E}^{2} \sim 1 $. Due to our assumption $\alpha T>>1$ 
(with $T=T_{P}\sim\frac{\eta}{E^{2}}$) we get the condition 
\begin{equation}
{\cal E}^{2} \sim \frac{E^{2}}{\eta}<<\alpha.
\end{equation}
Now, using the fact that 
$\frac{E^{2}}{\alpha\eta}<<1<<\frac{E^{2}}{\alpha\eta v}$ 
- which follows from the pole equation 
and the nonrelativistic limit -
we end up with:
\begin{equation}
v<<\frac{{\cal E}^{2}}{\alpha}<<1.
\end{equation}
Other relationships are possible among our parameters.
For instance the relation  
$\alpha v T<<1<<\alpha T$  could make us infer that 
$\alpha  T \sqrt{v}\sim 1$
which is a reasonable assumption.  
These constraints can be further refined if 
we assume a definite relationship 
between $\alpha$ and $\beta$ ($\alpha \sim \beta$, or
$\alpha \sqrt{v} \sim \beta$, for instance). 
However, we prefer not to add any 
further assumption for the time being. 
We just stress again that in order for the 
pair creation of vortices in a thin 
superconductor to be observable we need (30) to be satisfied.
We postpone further elaborations for the time 
when experimental results will be in sight.  

\vskip1cm

{\bf Aknowledgements}: We thank Claudio Scrucca for useful discussions. 

\vskip1.5cm

\centerline{\large \bf Appendix A}

\vskip0.5cm

We review here the boundary state formalism \cite{callan},
applying it to our case. 
The strategy is the following: We first
establish the boundary conditions along $\sigma$, 
in the open string channel. 
We then switch to the closed string
channel and obtain the operatorial boundary 
conditions there. Our amplitude 
will be the scalar product between the state 
satisfying them at 
$\sigma =0$ and  at $\sigma =l$. Since 
the uncoupled case ($X_{2}$) 
is easily obtainable
from the coupled one, we restrict to the 
action along $X_{0}$ and $X_{1}$: 

\begin{displaymath}
S=\int_{0}^{t} d\tau \int^{l}_{0}d\sigma [-\frac{\alpha}{2}
(\frac{\partial X^{0}}{\partial \tau})^{2}-\frac{\alpha}{2}
(\frac{\partial X^{0}}{\partial \sigma})^{2}+\frac{\beta}{2}
 (\frac{\partial X^{1}}{\partial \tau})^{2}+\frac{\beta v^{2}}{2}
(\frac{\partial X^{1}}{\partial \sigma})^{2}]
-i\int_{0}^{t} d\tau [EX^{0}\frac{\partial X^{1}}{\partial \tau}]_{\sigma =l}
\end{displaymath}
The boundary conditions at $\sigma=l$ are:
\begin{equation}
(\alpha \frac{\partial X_{n}^{0}}{\partial\sigma}
+iE\frac{\partial X_{n}^{1}}{\partial \tau})_{\sigma=l}=0
\end{equation}
\begin{equation}
(\beta v^{2}\frac{\partial X_{n}^{1}}{\partial\sigma}
+iE\frac{\partial X_{n}^{0}}{\partial \tau})_{\sigma=l}=0.
\end{equation}
At $\sigma=0$ we just put $E=0$.
Assuming now periodic boundary 
conditions along $\tau$,
$ X(\tau,\sigma )=X(t+\tau ,\sigma )$,
we Fourier expand $X$ with respect to $\tau$ :
\begin{displaymath}
X(\tau,\sigma)=\sum_{nZ}e^{i\omega_{n} \tau}X_{n}(\sigma) 
\quad {\rm where} \quad
\omega_{n}=\frac{2\pi}{t}n.
\end{displaymath}
The boundary conditions for the $X_{n}$'s are given by the 
Fourier transform of eqs.(32-33):
\begin{equation}
\frac{\partial X_{n}^{0}}{\partial\sigma}|_{\sigma=0}=0 \quad \quad   
     \alpha\frac{\partial X_{n}^{0}}{\partial\sigma}\mid_{\sigma=l}
-\omega_{n}EX_{n}^{1}\mid_{\sigma=l}=0   
\end{equation}
\begin{equation}
\frac{\partial X_{n}^{1}}{\partial\sigma}\mid_{\sigma=0}=0 \quad \quad
    \beta v^{2}\frac{\partial X_{n}^{1}}{\partial\sigma}\mid_{\sigma=l}
-\omega_{n}EX_{n}^{0}\mid_{\sigma=l}=0 .  
\end{equation}
The free equations of motion (there is no electric field in the bulk) are:
$$
\frac{\delta S(n)}{\delta X_{-n}^{0}(\sigma)} 
= -\alpha \omega_{n}^{2}
X_{n}^{0}+\alpha\frac{\partial^{2}X_{n}^{0}}{\partial\sigma^{2}}=0 
$$
$$
\frac{\delta S(n)}{\delta X_{-n}^{1}(\sigma)}=\beta\omega_{n}^{2}X_{n}^{1}
-\beta v^{2}\frac{\partial^{2}X_{n}^{1}}{\partial\sigma^{2}}=0.   
$$

Until now, we spoke about the periodic propagation 
of an open string
and consequently developed X on open-string modes.
Now we reverse the picture (i.e. the roles of  
Euclidean $\tau$ and $\sigma$) and look 
at a  closed string parametrised by $\tau$, propagating 
in the 'time' $\sigma$, for $\sigma$ from 
$0$ to $l$. Hence let us develop X in closed-string modes: 

%\begin{equation}
%\begin{array}
\begin{displaymath}
 X(t,\sigma)  =X_{0}+\frac{i}{\sqrt{4\pi}} \sum_{n\geq 1}\frac{1}{\sqrt{n}}
(e^{i\omega_{n}\tau-\omega_{n}\sigma\frac{1}{v}}X_{n}-
e^{-i\omega_{n}\tau+\omega_{n}\sigma\frac{1}{v}}X_{-n}+ 
\end{displaymath}
\begin{equation}
 +\tilde{X}_{n}e^{-i\omega_{n}\tau-\omega_{n}\sigma \frac{1}{v}}-
\tilde{X_{-n}}e^{i\omega_{n}\tau+\omega_{n}\sigma\frac{1}{v}}).
\end{equation}
%\end{array}
%\end{equation}
Using that we  obtain the boundary 
conditions for the
closed string modes: 

\begin{equation}
X_{n}^{0}(l)+\tilde{X}_{-n}^{0}(l)=-\frac{E}{\alpha}
(X_{n}^{1}(l)-\tilde{X}_{-n}^{1}(l)) \quad \quad \forall n
\end{equation}

\begin{equation}
X_{n}^{1}(l)+\tilde{X}_{-n}^{1}(l)=-\frac{E}{\beta v}(X_{n}^{0}(l)-\tilde{X}_{-n}^{0}(l)) \quad \quad \forall n.
\end{equation}
At $\sigma=0$ the right-hand-side vanishes. For $E=0$
we get the boundary condition for an uncoupled coordinate $X$:
\begin{equation}
X_{n}(\sigma=0,l)+\tilde{X}_{-n}(\sigma=0,l)=0 \quad \quad \forall n.
\end{equation}
Next we find for the boundary state $|B_{E}>$ satisfying (37,38) 
the expression
(up to a normalization constant $N^{\frac{1}{2}}(E)$ ):

\begin{displaymath}
|B_{E}> =exp\left\{\frac{1}{\frac{E^{2}}{\alpha \beta v}-1}
\sum_{n>0}[(1+\frac{E^{2}}{\alpha\beta v}) (X_{-n}^{1}\tilde{X}_{-n}^{1} -X_{-n}^{0}\tilde{X}_{-n}^{0}) \right.
\end{displaymath}
\begin{equation}
  \left. +2E(\frac{X_{-n}^{0}\tilde{X}_{-n}^{1}}{\alpha}
 -\frac{X_{-n}^{1}\tilde{X}_{-n}^{0}}{\beta v})]\right\} |0> 
\end{equation} 
the state and the operators $X$ 
being taken at $\sigma=l$.
At $s=0$ this becomes 
$$|B_{0}>=exp\sum_{n \geq 1}(X_{-n}^{0} \tilde{X}_{-n}^{0}-X_{-n}^{1}\tilde{X}_{-n}^{1}) |0> .$$
Now, the quantity of interest is, up to a normalization factor $N(E)$:  
\begin{equation}
<B_E(l)|B_{0}>=<B_{E}|e^{-lH}|B_{0}>=N^{-1}(E) \int DX^{0}(\sigma ,t)
DX^{1}(\sigma ,t) e^{-S}
\end{equation}
which we are going to calculate 
in the main body of the paper, obtaining thus 
the required path integral.

\vskip1.5cm
\newpage 

\centerline{\large \bf Appendix B}

\vskip0.5cm

We prove here the equality - eq.(22) in the text:

\begin{equation}
<0|e^{\kappa b\tilde{b}}e^{\lambda a\tilde{a}}e^{\mu a\tilde{b}}e^{\nu b\tilde{a}}
e^{\rho a^{\dagger}\tilde{a}^{\dagger}}e^{\sigma b^{\dagger}\tilde{b}^{\dagger}}|0>
=\frac{1}{(1-\kappa\sigma)(1-\lambda\rho)-\mu\nu\rho\sigma}
\end{equation}
where $a$,$\tilde{a}$,$b$,$\tilde{b}$ are 
independent annihilation operators, while 
$a^{\dagger}$,$\tilde{a}^{\dagger}$,$b^{\dagger}$,$\tilde{b}^{\dagger}$
are the corresponding creation operators.

We expand the exponentials in power series 
in the left-hand-side of (42) and keep only 
the non-zero terms. Then, by using the harmonic oscillator
normalization $<0|a^{n}(a^{\dagger})^{n}|0>=n!$, as well as a trick - 
differentiating and then resuming - we obtain what follows:

\begin{eqnarray*}
\lefteqn{<0|e^{\kappa b\tilde{b}}e^{\lambda a\tilde{a}}e^{\mu a\tilde{b}}e^{\nu b\tilde{a}}
e^{\rho a^{\dagger}\tilde{a}^{\dagger}}e^{\sigma b^{\dagger}\tilde{b}^{\dagger}}|0>=} \\
& &   \\
& & =\sum_{m=0}^{\infty}\sum_{r=0}^{\infty}\sum_{s=0}^{\infty}
<0|\frac{(\kappa b\tilde{b})^{s}}{s!}\frac{(\lambda a\tilde{a})^{r}}{r!}\frac{(\mu a\tilde{b})^{m}}{m!}\frac{(\nu b\tilde{a})^{m}}{m!}
\frac{(\rho a^{\dagger}\tilde{a}^{\dagger})^{m+r}}{(m+r)!}\frac{(\sigma b^{\dagger}\tilde{b}^{\dagger})^{m+s}}{(m+s)!}|0>=  \\
& &    \\
& & =\sum_{m=0}^{\infty}\sum_{r=0}^{\infty}\sum_{s=0}^{\infty}
\frac{\kappa^{s}}{s!}\frac{\lambda^{r}}{r!}\frac{\mu^{m}}{m!}\frac{\nu^{m}}{m!}
\frac{\rho^{m+r}}{(m+r)!}\frac{\sigma^{m+s}}{(m+s)!} [(r+m)!(s+m)!]^{2}=   \\
& &    \\
& & =\sum_{m=0}^{\infty} \frac{1}{(m!)^{2}}(\mu\nu)^{m}\frac{d^{m}}{d\kappa^{m}}(\frac{1}{1-\kappa\sigma})\frac{d^{m}}{d\lambda^{m}}(\frac{1}{1-\lambda\rho})= \\
& & \\
& & =\sum_{m=0}^{\infty} (\mu\nu)^{m} (\rho\sigma)^{m} \frac{1}{(1-\kappa\sigma)^{m+1}} \frac{1}{(1-\lambda\rho)^{m+1}}= \\
& & \\
& & =\frac{1}{ (1-\kappa\sigma)(1-\lambda\rho)-\mu\nu\rho\sigma }. 
\end{eqnarray*}
For $\mu=\nu=0$ we also obtain as a particular case:
$$
<0|e^{\kappa b\tilde{b}}e^{\lambda a\tilde{a}}e^{\rho a^{\dagger}\tilde{a}^{\dagger}}e^{\sigma b^{\dagger}\tilde{b}^{\dagger}}|0>
=\frac{1}{ (1-\kappa\sigma)(1-\lambda\rho)}.
$$

\vskip1.5cm

\newpage

%\centerline{\large \bf References}

%\vskip0.5cm

\end{document}